\documentclass{actasen}

\begin{document}

%%ÉèÖÃÊ×Ò³Ò³Âë
\setcounter{page}{55}

\Volume{2015}{56}% Äê¡¢¾í

%%ҳüÉèÖÃ

\runheading{SHEN Hong}%

\title{A Complete Equation of State for Astrophysical Simulations}

\footnotetext{$^{\dag}$ Supported by the National Natural Science Foundation
of China (Grant No. 11375089)

$^{\bigtriangleup}$ shennankai@gmail.com\\
}

\enauthor{SHEN Hong$^{\bigtriangleup}$ }
{School of Physics, Nankai University, Tianjin 300071, China}

\abstract{
We construct a complete equation of state (EOS) covering a wide range of
temperature, proton fraction, and baryon density for the use of astrophysical
simulations. We employ the relativistic mean-field (RMF) theory to describe
nuclear interactions and adopt the Thomas-Fermi approximation to describe the
nonuniform nuclear matter. The uniform matter and nonuniform matter are studied
consistently using the same RMF theory.}


\maketitle

\section{Introduction}

The equation of state (EOS) is an important input in various astrophysical
studies like neutron stars and supernovae\rf{1}.
In past decades, many efforts have been devoted to construct the EOS for
supernova simulations\rf{2,3,4}.
There are two standard EOS's, which are commonly used in supernova simulations,
namely the one by Lattimer and Swesty\rf{2}
and the one by Shen et al.\rf{3}.
The Lattimer-Swesty EOS was made by using a compressible liquid-drop model
with Skyrme forces. The Shen EOS was calculated with the relativistic mean-field
(RMF) model and the Thomas-Fermi approximation.
In our previous work\rf{3}, we constructed the EOS table
covering a wide range of temperature $T$, proton fraction $Y_p$, and baryon mass
density $\rho_B$ for the use of supernova simulations.
Recently, we have recalculated the Shen EOS with an improved design of
ranges and grids according to the requirements of the EOS users,
and furthermore, the presence of hyperons has also been considered\rf{5}.

\section{Method}

We employ the RMF theory to describe nuclear matter with uniform or nonuniform distributions. The RMF theory has been successfully used to study various
phenomena in nuclear physics.
We adopt the TM1 parameter set, which can provide a good description of nuclear matter and finite nuclei\rf{6}.
The Lagrangian density in the RMF theory with the TM1 parameter set is given by
\begin{eqnarray}
\label{eq:LRMF}
{\cal L}_{\rm{RMF}} = & \bar{\psi}\left[i\gamma_{\mu}\partial^{\mu} -M
-g_{\sigma}\sigma-g_{\omega}\gamma_{\mu}\omega^{\mu}
-g_{\rho}\gamma_{\mu}\tau_a\rho^{a\mu}
\right]\psi
 +\frac{1}{2}\partial_{\mu}\sigma\partial^{\mu}\sigma
-\frac{1}{2}m^2_{\sigma}\sigma^2-\frac{1}{3}g_{2}\sigma^{3}
 \nonumber\\
 & -\frac{1}{4}g_{3}\sigma^{4}
   -\frac{1}{4}W_{\mu\nu}W^{\mu\nu}
+\frac{1}{2}m^2_{\omega}\omega_{\mu}\omega^{\mu}
+\frac{1}{4}c_{3}\left(\omega_{\mu}\omega^{\mu}\right)^2
-\frac{1}{4}R^a_{\mu\nu}R^{a\mu\nu}
+\frac{1}{2}m^2_{\rho}\rho^a_{\mu}\rho^{a\mu} ,
\end{eqnarray}
where $W^{\mu\nu}$ and $R^{a\mu\nu}$ are the antisymmetric field tensors
for $\omega^{\mu}$ and  $\rho^{a\mu}$, respectively.
In the RMF approach, meson fields are treated as classical fields
and the field operators are replaced by their expectation values.
Starting from the Lagrangian density, we can derive the equations of motion
for nucleons and mesons, which are then solved self-consistently.

We study properties of dense matter with both uniform and nonuniform
distributions. For uniform matter, the RMF theory can be easily used
to calculate properties of nuclear matter. For nonuniform matter,
we assume each heavy nucleus is
located at the center of a charge-neutral
Wigner--Seitz cell consisting of a vapor of nucleons, electrons, and alpha-particles. The nucleon distribution in the Wigner--Seitz cell, $n_i(r)$ ($i=p$ or $n$),
is assumed to have the form
\begin{equation}
\label{eq:nitf}
n_i\left(r\right)=\left\{
\begin{array}{ll}
\left(n_i^{\rm{in}}-n_i^{\rm{out}}\right) \left[1-\left(\frac{r}{R_i}\right)^{t_i}
\right]^3 +n_i^{\rm{out}},  & 0 \leq r \leq R_i, \\
n_i^{\rm{out}},  & R_i \leq r \leq R_{\rm{cell}}, \\
\end{array} \right.
\end{equation}
where $R_i$ and $R_{\rm{cell}}$ represent the radii of the nucleus
and the Wigner--Seitz cell, respectively.
The density parameters $n_i^{\rm{in}}$ and $n_i^{\rm{out}}$ are the densities at
$r=0$ and $ r \geq R_i $, while $t_i$ determines the relative surface thickness
of the nucleus. The distribution of alpha-particles is assumed to have a similar
form, which decreases as $r$ approaches the center of the nucleus.
For a system with fixed temperature $T$, proton fraction $Y_p$, and baryon mass
density $\rho_B$, the thermodynamically favorable state is determined by minimizing
the free energy density with respect to these parameters.
The results of the RMF model are used in the Thomas-Fermi calculation,
so the treatments of nonuniform and uniform matter are consistent.

\section{Results}

At each given $T$, $Y_p$, and $\rho_B$, we perform the minimization of the free
energy. The thermodynamically favorable state is the one having the lowest
free energy. By comparing the free energies of nonuniform matter and uniform matter,
we can determine the most favorable state and examine the phase transition between
nonuniform matter and uniform matter.

In Fig. 1, we show the phase diagram in the $\rho_B$--$T$ plane for $Y_p=0.3$.
The shaded region corresponds to the nonuniform matter phase, in which heavy
nuclei are formed to lower the free energy of the system.
The dashed line denotes the boundary where the alpha-particle fraction
$X_{\alpha}$ changes between $X_{\alpha}<10^{-4}$ and $X_{\alpha}>10^{-4}$.
It is evident that heavy nuclei only exist in the medium-density and low-temperature region.
\begin{figure}[tbph]
\begin{minipage}{0.6\textwidth}
\includegraphics[bb=50 315 525 595,width=7.0cm]{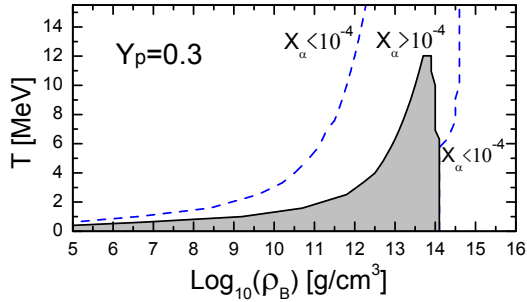}
\end{minipage}%
\begin{minipage}{0.4\textwidth}
\caption{Phase diagram of nuclear matter in the $\rho_B$--$T$ plane for $Y_p=0.3$.}
\end{minipage}
\end{figure}

In Table 1, we compare the improved EOS's, namely EOS2 and EOS3\rf{5}, with the previous EOS1\rf{3}. The difference between
EOS2 and EOS3 is that only the nucleon degree of freedom is
considered in EOS2, while EOS3 includes additional contributions from
$\Lambda$ hyperons. In comparison with EOS1, several
improvements have been made in EOS2 and EOS3 according
to the requirements of the users.
The grid spacing for temperature $T$ is significantly reduced
and a linear $Y_p$ grid is adopted instead of the logarithmic $Y_p$
grid used in EOS1. These improvements are very important for numerical
simulations of supernovae.
\begin{table}[htb]
\scriptsize
\caption{Comparison between the EOS's discussed in this paper
\label{tab:1}}
\begin{center}
\begin{tabular}{ccccc}
\hline\hline
 &  & EOS1 & EOS2 & EOS3 \\
\hline
 Constituents &  & $n$, $p$, $\alpha$      & $n$, $p$, $\alpha$      & $n$, $p$, $\alpha$, $\Lambda$ \\
\hline
 $T$ & Range            & $ -1.0 \leq \log_{10}(T) \leq 2.0 $
                        & $ -1.0 \leq \log_{10}(T) \leq 2.6 $
                        & $ -1.0 \leq \log_{10}(T) \leq 2.6 $ \\
(MeV) & Grid spacing    & $\Delta \log_{10}(T)\simeq 0.1$
                        & $\Delta \log_{10}(T)=0.04$
                        & $\Delta \log_{10}(T)=0.04$ \\
\hline
 $Y_p$ & Range          & $ -2 \leq \log_{10}(Y_p) \leq -0.25 $
                        & $ 0 \leq Y_p \leq 0.65 $
                        & $ 0 \leq Y_p \leq 0.65 $   \\
       & Grid spacing   & $\Delta \log_{10}(Y_p)=0.025$
                        & $\Delta Y_p=0.01$
                        & $\Delta Y_p=0.01$ \\
\hline
 $\rho_B$ & Range       & $  5.1 \leq \log_{10}(\rho_B) \leq 15.4 $
                        & $  5.1 \leq \log_{10}(\rho_B) \leq 16 $
                        & $  5.1 \leq \log_{10}(\rho_B) \leq 16 $  \\
 $\rm{(g\,cm^{-3})}$ & Grid spacing & $\Delta \log_{10}(\rho_B)\simeq 0.1 $
                                & $\Delta \log_{10}(\rho_B) = 0.1 $
                                & $\Delta \log_{10}(\rho_B) = 0.1 $ \\
      \hline\hline
\end{tabular}
\end{center}
\end{table}


\begin{thebibliography}{999}

\bibitem{1}
Janka~H Th, Langanke K, Marek A, et al.
Phys. Rep., 2007, 442, 38

\bibitem{2}
Lattimer J M, Swesty F D.
Nucl. Phys. A, 1991, 535, 331

\bibitem{3}
Shen H, Toki H, Oyamatsu K, Sumiyoshi K.
Prog. Theor. Phys., 1998, 100, 1013

\bibitem{4}
Shen G, Horowitz C J, Teige S.
Phys. Rev. C, 2010, 82, 015806

\bibitem{5}
Shen H, Toki H, Oyamatsu K, Sumiyoshi K.
Astrophys. J. Suppl., 2011, 197, 20

\bibitem{6}
Sugahara Y, Toki H.
Nucl. Phys. A, 1994, 579, 557

\end{thebibliography}
\end{document}